\documentclass[12pt,a4paper,final]{iopart}
\usepackage{iopams}
\usepackage{iopams}  
\usepackage{graphicx}
\usepackage[breaklinks=true,colorlinks=true,linkcolor=blue,urlcolor=blue,citecolor=blue]{hyperref}
\usepackage[usenames,dvipsnames,svgnames,table]{xcolor}

\begin{document}

\title[Electronic and magnetic properties of superconducting \textit{Ln}O$_{1-x}$F$_{x}$BiS$_{2}$]{Electronic and magnetic properties of superconducting \textit{Ln}O$_{1-x}$F$_{x}$BiS$_{2}$ (\textit{Ln} = La, Ce, Pr, and Nd) from first principles}
\author{Corentin Morice}
\ead{cm712@cam.ac.uk}
\address{Cavendish Laboratory, University of Cambridge, Cambridge CB3 0HE, United Kingdom}
\author{Emilio Artacho}
\ead{ea245@cam.ac.uk}
\address{Cavendish Laboratory, University of Cambridge, Cambridge CB3 0HE, United Kingdom}
\address{Nanogune and DIPC, Tolosa Hiribidea 76, 20018 San Sebasti\'an, Spain}
\address{Basque Foundation for Science, Ikerbasque, 48011 Bilbao, Spain}
\author{Si\^an E. Dutton}
\address{Cavendish Laboratory, University of Cambridge, Cambridge CB3 0HE, United Kingdom}
\author{Hyeong-Jin Kim}
\address{Cavendish Laboratory, University of Cambridge, Cambridge CB3 0HE, United Kingdom}
\author{Siddharth S. Saxena}
\ead{sss21@cam.ac.uk}
\address{Cavendish Laboratory, University of Cambridge, Cambridge CB3 0HE, United Kingdom}
\date{\today}

\begin{abstract}
A density functional theory study of the BiS$_{2}$ superconductors containing rare-earths: \textit{Ln}O$_{1-x}$F$_{x}$BiS$_{2}$ (\textit{Ln} = La, Ce, Pr, and Nd) is presented. We find that CeO$_{0.5}$F$_{0.5}$BiS$_{2}$ has competing ferromagnetic and weak antiferromagnetic tendencies, the first one corresponding to experimental results. We show that PrO$_{0.5}$F$_{0.5}$BiS$_{2}$ has a strong tendency for magnetic order, which can be ferromagnetic or antiferromagnetic depending on subtle differences in 4$f$ orbital occupations. We demonstrate that NdO$_{0.5}$F$_{0.5}$BiS$_{2}$ has a stable magnetic ground state with weak tendency to order. Finally, we show that the change of rare earth does not affect the Fermi surface, and predict that CeOBiS$_{2}$ should display a pressure induced phase transition to a metallic, if not superconducting, phase under pressure.

\end{abstract}

\maketitle

\section{Introduction}

The excitement generated by the recent finding of two new superconductors, Bi$_{3}$O$_{2}$S$_{3}$ and LaO$_{1-x}$F$_{x}$BiS$_{2}$ \cite{Mizuguchi2012a,Mizuguchi2012}, led to the unveiling of a whole class of superconductors containing BiS$_{2}$ bilayers. These compounds share a common two-dimensional structure with alternating BiS$_{2}$ bilayers and spacer layers. It is the BiS planes in the BiS$_{2}$ bilayers which are thought to be responsible for superconductivity in these compounds \cite{Morice2015}. Soon after the discovery of superconductivity in LaO$_{1-x}$F$_{x}$BiS$_{2}$, chemical substitution to tune the properties was attempted. The first research axis was to replace the lanthanum atom by other lanthanides: cerium, praseodymium, neodymium, europium and ytterbium \cite{Demura2013,Jha2013a,Jha2013,Xing2012,Suzuki2015,Guo2015,Zhai2014,Goto2015,Yazici2013}. Hole and electron doping of the parent compound has also been studied in Sr$_{x}$La$_{1-x}$FBiS$_{2}$ \cite{Lin2013} and La$_{1-x}$\textit{M}$_{x}$OBiS$_{2}$ (\textit{M} = Ti, Zr, Hf, Th) \cite{Yazici2013a,Benayad2014} respectively.

The electronic and thermodynamic properties of this group of materials are indeed interesting and theoretical suggestions for the mechanism of superconductivity range from spin-fluctuation mediated superconductivity \cite{Martins2013}, to proximity to ferroelectricity and charge density wave (CDW) instabilities \cite{Yildirim2013,Wan2013}. Recent work even revealed the potential of related materials for photovoltaic applications \cite{Meng2015}. Ferromagnetism has been shown experimentally to coexist with superconductivity in CeO$_{0.5}$F$_{0.5}$BiS$_{2}$ \cite{Jha2013b, Demura2015, Kajitani2015, Lee2014, Miura2015, Paris2014, Sugimoto2014, Xing2012, Sugimoto2015}, and quantum critical fluctuations of the magnetic moments have been observed in CeOBiS$_{2}$ \cite{Higashinaka2015}.

The electronic structure of the two first compounds to be found, Bi$_{3}$O$_{2}$S$_{3}$ \cite{Morice2015} and La(O,F)BiS$_{2}$ \cite{Shein2013,Usui2012,Suzuki201321,Yildirim2013}, was calculated. These calculations indicated that the superconducting electrons are a mixture of Bismuth 6\textit{p$_{x,y}$} and Sulphur 3\textit{p$_{x,y}$} states \cite{Shein2013,Usui2012}. These form eight bands, four of which are under the Fermi level, when the other four either are above the Fermi level or cross it.

The coupling mechanism for superconductivity has been investigated in various ways. Electron-phonon interactions have been calculated in La(O,F)BiS$_{2}$, and yield a large electron-phonon coupling constant, suggesting superconductivity in this compound is strongly coupled and conventional \cite{Wan2013, Yildirim2013, Li2013}.  Renormalisation-group calculations suggested triplet pairing and weak topological superconductivity \cite{Yao2013,Yang2013}, a possibility studied in the context of quasiparticle interference \cite{Gao2014}. Random phase approximation was applied to a two-orbital model \cite{Usui2012}, leading to an extended s-wave or d-wave pairing \cite{Martins2013,Zhou2013}.

\begin{figure}
\center{\includegraphics[scale=0.8]{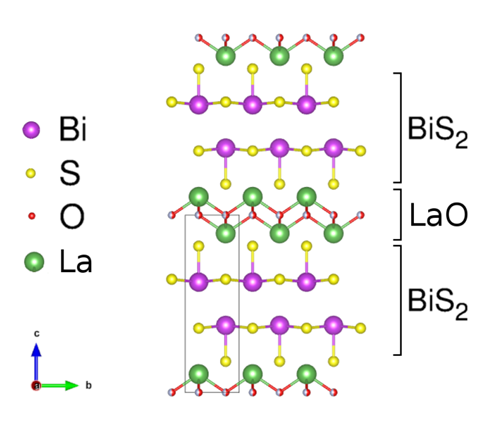}
\caption{Crystal structure of LaOBiS$_{2}$, represented in VESTA \cite{Momma2011}. All the compounds studied share this same structure, with replacement of lanthanum by other lanthanides, or of oxygen by fluorine. The black solid line represents the unit cell, which contains two formula units.}
\label{CrystalStructure}}
\end{figure}

In this paper we present our findings from electronic structure calculations for the stoichiometric and doped series of materials \textit{Ln}O$_{1-x}$F$_{x}$BiS$_{2}$ with \textit{Ln} = La, Ce, Pr and Nd, and $x=0,0.5$. The doped phases calculated have been reported as being superconducting, and the parent phases have been reported as being non-superconducting.

We first show that the change of rare earth does not affect the Fermi surface and that the conduction electrons are confined to the bismuth-sulphur planes. We then demonstrate that CeO$_{0.5}$F$_{0.5}$BiS$_{2}$ has competing ferromagnetic and weak antiferromagnetic tendencies. The ferromagnetic tendency is in agreement with experimental results. We show that PrO$_{0.5}$F$_{0.5}$BiS$_{2}$ has a strong tendency to magnetic order, which could be ferromagnetic or antiferromagnetic depending on subtle differences in 4$f$ orbital occupations. We show that NdO$_{0.5}$F$_{0.5}$BiS$_{2}$ has a stable weakly antiferromagnetic ground state. Finally, we predict that applying pressure to CeOBiS$_{2}$ would make it metallic, and possibly superconducting.

\section{Methods}

We limit our study to materials which crystal structure has been experimentally determined, with the exception of the Ce doped compound. We have been able to provide a structure through relaxation methods for the Ce compound, since the availability of the La, Pr and Nd compounds provides a reasonable reliability on the starting point.

Band structures and densities of states were calculated within density functional theory using the SIESTA method \cite{Artacho2008,Soler2002}, implementing the generalized gradient approximation (GGA) in the shape of the Perdew, Burke, and Ernzerhof functional \cite{Perdew1996} and the GGA+$U$ method. It uses norm-conserving pseudopotentials to replace the core electrons, while the valence electrons are described using atomic-like orbitals as basis states at the double zeta polarized level. These pseudopotentials include scalar relativistic corrections. Spin-orbit coupling was not included but has been shown in previous work to be of no relevance to the main features at the Fermi level in this family of materials \cite{Morice2015,Wan2013}. Details on the basis sets used are available in the supplemental material.

Calculations for $x=0.5$ have been performed for all the compounds. All these structures share the same space group: P4/nmm (Figure \ref{CrystalStructure}). The calculations for the lanthanum compound were performed without spin polarisation. However for \textit{Ln} $\ne$ La, the presence of unpaired 4$f$ electrons demands the use of spin polarisation in the calculations. Calculations for the La compound were performed using experimental data for the geometry obtained after high pressure treatment of the sample, for which the maximal superconducting $T_{c}$ (10.6 K) was found \cite{Mizuguchi2012a}. For the Pr and Nd compounds, we used structural experimental data for the systems for which $T_{c}$ of 5.5 K \cite{Jha2013a} and 5 K were found \cite{Jha2013} respectively.

No atomic coordinates have been reported for the Ce compound yet, so we obtained a geometry from Density Functional Theory (DFT) by relaxing the structure (both atoms and cell), starting from the structure known for the La compound. Two relaxations were performed, one with a $k$-grid of 8x8x3 points, a standard set of parameters and without $U$, and one with a $k$-grid of 15x15x5 points, a high quality set of parameters and with $+U$. Both relaxations were performed in the ferromagnetic configuration. The comparison of the results shows the relaxation is well converged: the lattice constants differ by approximately 2 m\AA , and the fractional atomic coordinates by less than $0.0004$. The details of the crystal structures are available in the supplementary material. The obtained atomic positions and $a$ lattice parameter are very close to the ones of the Pr doped compound. The $c$ lattice parameter is significantly smaller than the ones of the other doped compounds. This is in accordance with mismatches between DFT and experiments in the inter-layer forces noted in other compounds of this family of superconductors \cite{Morice2015}.

There are two O/F sites in each unit cell of every doped compound. It has been shown that disorder in the O and F positions has a very minor effect on the electronic structure at the level discussed in this work \cite{Lee2013}. As already done in \cite{Yildirim2013, Shein2013, Suzuki201321}, we assign O to (0,0,0) and F to (0.5,0.5,0) in the La doped compound and O to (0.75,0.25,0) and F to (0.25,0.75,0) in the other doped compounds. This changes the space group of the crystal structures of the doped compounds to P\={4}m2.

\begin{figure}
\center{\includegraphics[scale=0.3]{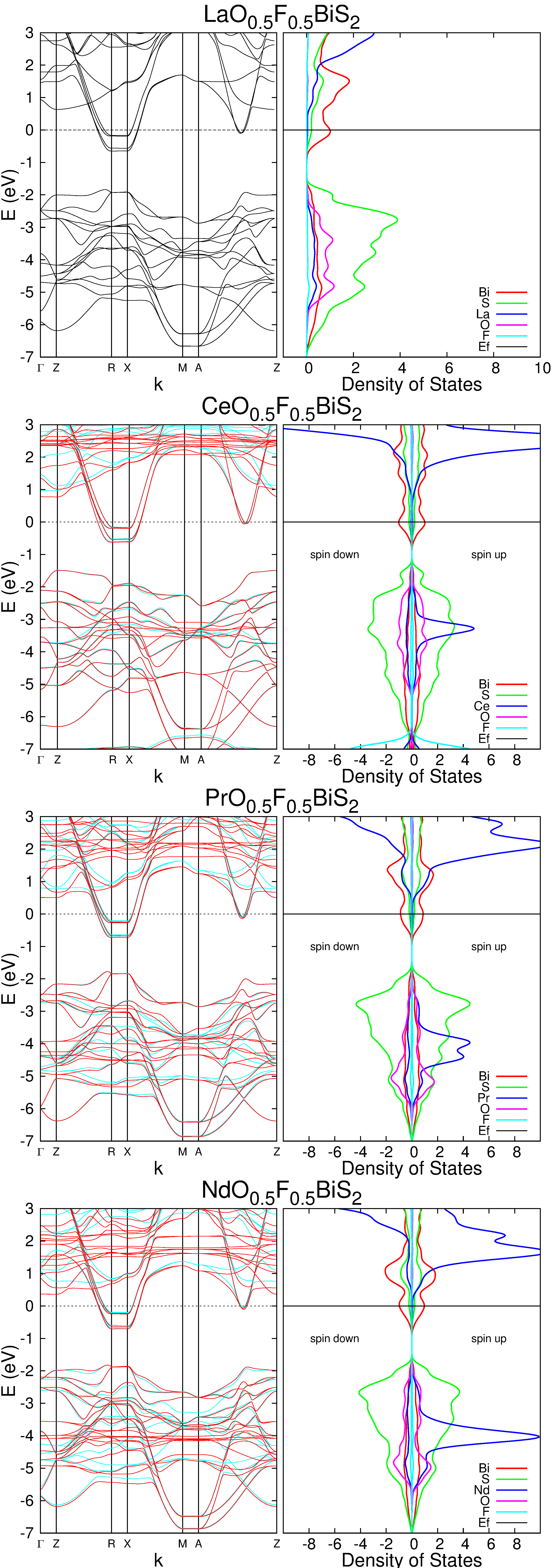}
\caption{Band structures (left) and densities of states projected onto the basis orbitals for Bi (red), S (green), \textit{Ln} (blue), O (lilac), and F (light blue) (right) for the four doped compounds. For the three compositions having 4$f$ electrons, we use $U=5$. The lanthanum compound band structure is non spin polarised, whereas the other three are spin polarised, in the ferromagnetic configuration.}
\label{BandStructures}}
\end{figure}

We also performed calculations on two parent phases: LaOBiS$_{2}$ and CeOBiS$_{2}$. Their space group is also P4/nmm. We used experimental structural data for both \cite{Tanryverdiev1995,Ceolin1976}, and performed spin-polarised calculations in the ferromagnetic configuration for CeOBiS$_{2}$. There are no experimental data available for the other two parent compounds to our knowledge.

We performed GGA+$U$ calculations \cite{Anisimov1991} for all the compounds except the lanthanum ones, because of the known strong correlation of the 4$f$ electrons in cerium, praseodymium and neodymium. The current implementation of DFT+$U$ in SIESTA is based on the formulation by Dudarev et al. \cite{Dudarev1998} which combines the two parameters $U$ and $J$ to produce an effective Coulomb repulsion $U_{eff} =U-J$ which we call $U$ in the following. The double-counting scheme used is also the one described by Dudarev et al. \cite{Dudarev1998}. Given the intrinsic difficulty in the ab-initio determination of a value of $U$ \cite{Loschen2007}, we explored the behaviour of the system for $U$ varying from 3 to 7 eV. The influence of $U$ on relaxations was investigated and proven to be negligible. Here we discuss results obtained with $U=5$ eV, given that this value is standard for the atoms we are dealing with \cite{Jiang2009,Yeriskin2010}. We tested the robustness of the results for $U$ between 3 eV and 7 eV.

Monkhorst-Pack sampling of the Brillouin zone was used for all calculations. $k$-point sampling proved to be quite delicate, and large sets were used to obtain reasonably converged results, especially for the convergence of magnetic results. The grids used were 25x25x8 for the Ce and Nd compounds, 35x35x11 for the Pr compound and 24x24x5 for the La compound. The band structure results converged for lower sets, and the results in Figure \ref{MagnetismU} were obtained for 8x8x3 sets.

\section{Results}

\subsection{Band structures}

\begin{figure}
\center{\includegraphics[scale=0.6]{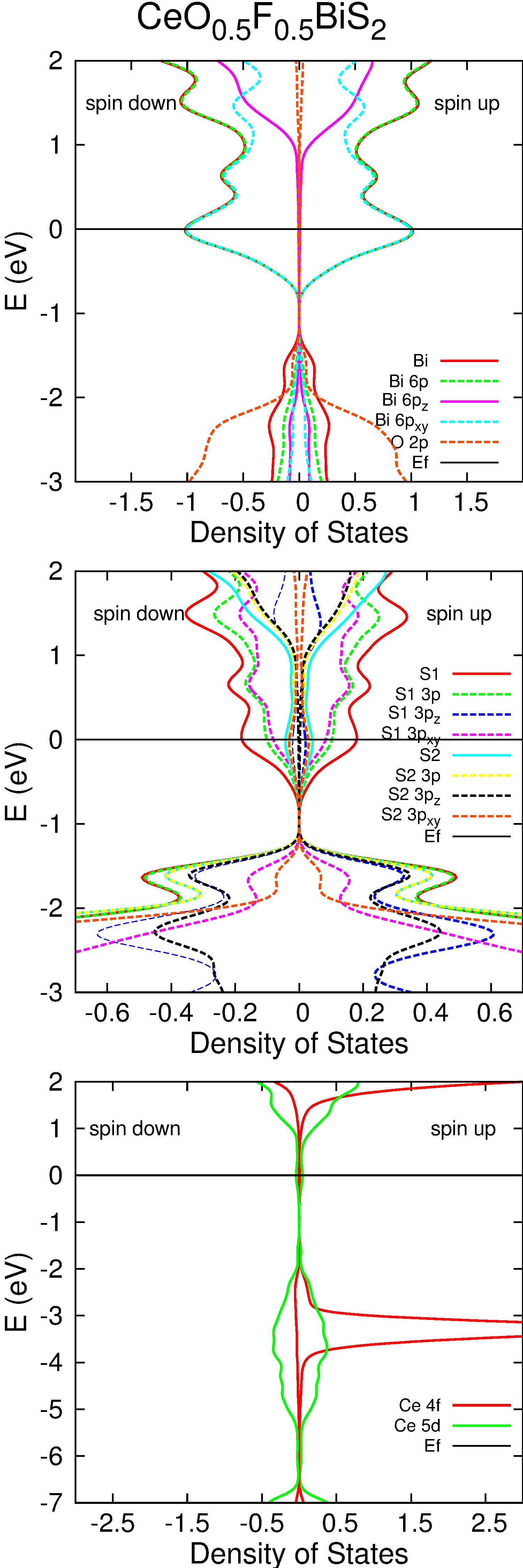}
\caption{Densities of states in CeO$_{0.5}$F$_{0.5}$BiS$_{2}$, projected onto several basis orbitals. S1 stands for the in-plane sulphur atom, and S2 stands for the out-of-plane sulphur atom. The label $p_{xy}$ is a shorthand for $p_{x}$ and $p_{y}$ orbitals which have the same density of states. We use $U=5$ and spin polarisation.}
\label{BiSbands}}
\end{figure}

We calculated the band structure of the four doped compounds. We observe many similarities between them, due to their proximity in terms of composition and crystal structure (Figure \ref{BandStructures}). More precisely, we find the same four bands (or eight bands with spin polarisation) near the R and X points, and two (or four) bands near the middle of the A-Z segment close to the Fermi level in all the compounds. They are similar to the ones in the Bi-O-S compounds \cite{Morice2015}. They are crossing the Fermi level for the doped compounds, and lie above the Fermi level for the two parent phases. The projected densities of states indicate that these bands are formed from Bi and S states. It has already been shown in Bi$_{3}$O$_{2}$S$_{3}$ and in the La compound that the main contribution to these bands comes from Bi 6$p$ states and S 3$p$ states. It is confirmed in all the systems considered here, as Figure \ref{BiSbands} shows. There we plotted the results for CeO$_{0.5}$F$_{0.5}$BiS$_{2}$, these being similar in all the compounds considered. The position of the Fermi level with respect to the bottom of these bands is very similar in all the compounds with $x=0.5$. The main effect of the fluorine doping is a shift of the Fermi level, as reported previously \cite{Shein2013,Usui2012,Suzuki201321}.

\begin{figure}
\center{\includegraphics[scale=0.25]{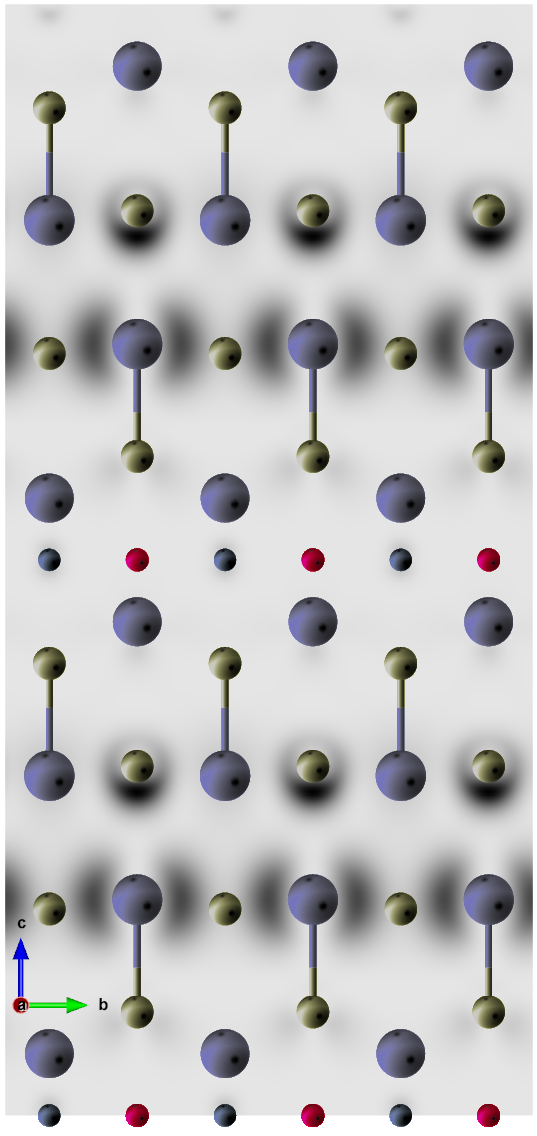}
\caption{Local density of states of CeO$_{0.5}$F$_{0.5}$BiS$_{2}$ in real space integrated in the energy range [-0.5,0], plotted with XCrySDen \cite{Kokalj2003}. Half of the atoms represented are not in the plane corresponding to the density plot: the atoms in the unit cell are arranged on two planes parallel to the x-z plane. The density plot corresponds to only one of these two planes. The BiS planes (horizontal and perpendicular to the figure) gather the vast majority of the density.}
\label{LDOS}}
\end{figure}

To explore further the character of the bands just below the Fermi level, we plotted the local density of states (LDOS) resolved in real space for the $x=0.5$ compounds for energies integrated between -0.5 eV and 0 eV (Figure \ref{LDOS}). The results confirm what had been found in other BiS$_{2}$-based compound: all these electrons accumulate around the BiS planes \cite{Morice2015}. This can also be seen in the partial density of states at the Fermi level (Figure \ref{BiSbands}) which is very low for the out-of-plane sulphur atoms. This raises the question of the existence of superconductors that have BiS planes without BiS$_{2}$ layers. Some $p_z$ participation is involved in the slight puckering of the BiS planes, inducing a clear distortion (along z) of the LDOS.

The lanthanide bands are at least 3 eV below and 0.5 eV above the Fermi level. The Fermi surface is robust to lanthanide substitution and to the change of $U$ between 3 eV and 7 eV. There is no difference in the density of states at the Fermi energy between compounds, the lanthanides' bands being far from the Fermi level, as expected, because of the strong correlation of the 4$f$ electrons. The extra electrons remain in the lanthanide bands and away from the Fermi level, hence the change of lanthanide has almost no effect. The Fermi surface is almost identical in the four cases, and does not seem to be responsible for the substantial change observed in $T_{c}$, nor does the band structure as a whole.

\subsection{Magnetic order}

We considered the magnetic properties of the doped compounds. They each have two \textit{Ln} atoms per unit cell. We compared the total energies of the ferromagnetic configuration, where the spins of the two atoms are aligned, and of the only antiferromagnetic configuration available in one unit cell, where the spins of the two atoms are antialigned.

\begin{figure}
\center{\includegraphics[scale=0.3]{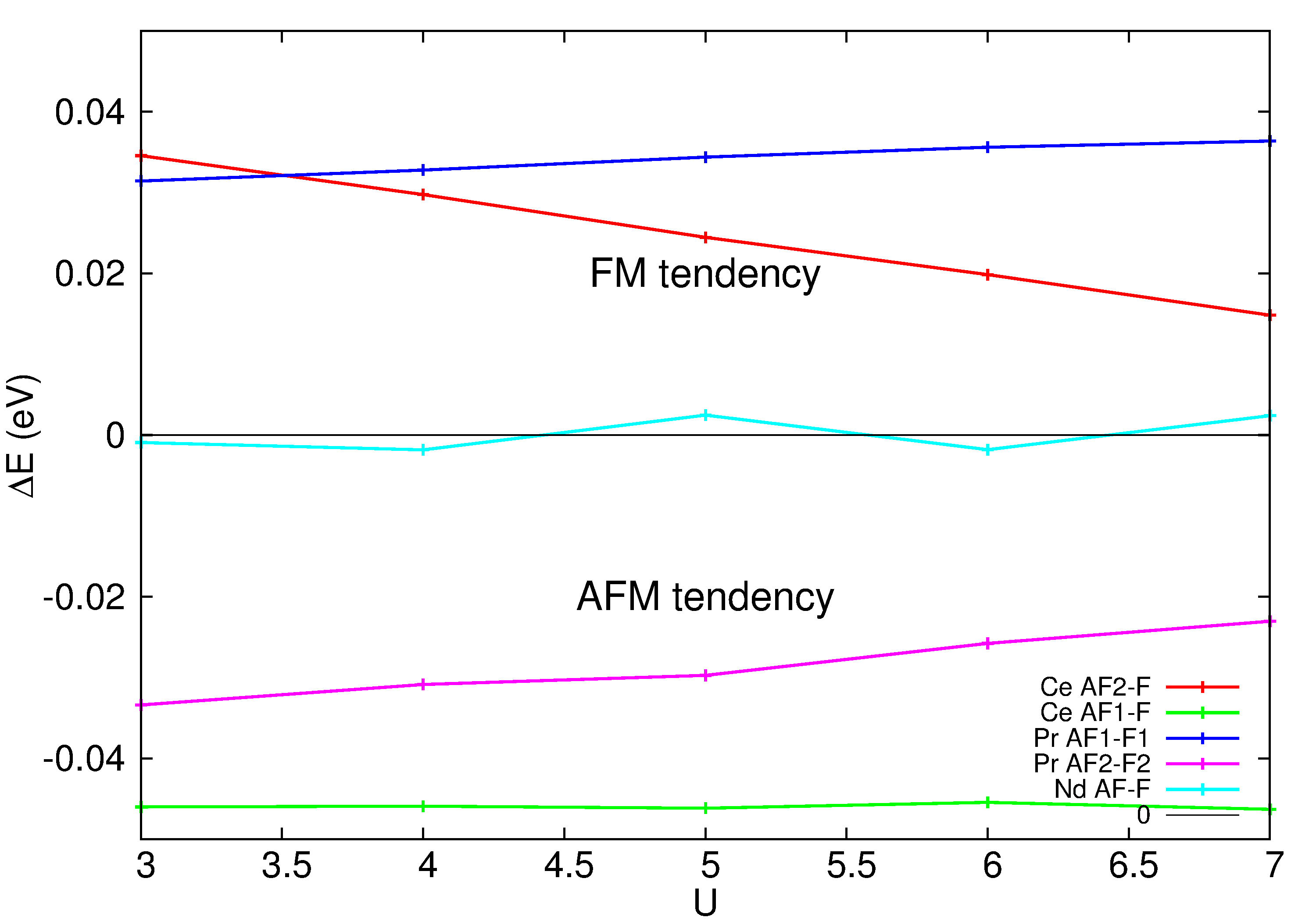}
\caption{Difference of total energy per unit cell of \textit{Ln}O$_{0.5}$F$_{0.5}$BiS$_{2}$ (\textit{Ln}=Ce, Pr, Nd)  between antiferromagnetic state and ferromagnetic states, for $U$ varying between 3 and 7. The $k$-point sampling set used here is smaller than in the rest of the text (see the Methods section). The key is labelled with names of magnetic configurations defined in the main text. A positive difference means the tendency is ferromagnetic.}
\label{MagnetismU}}
\end{figure}

Bulk cerium, praseodymium and neodymium have respectively one, three and four 4$f$ electrons per atom. The two-electrons difference between Ce and Pr is due to the fact that bulk cerium has one electron in a 5$d$ band, whereas bulk praseodymium and neodymium have not. However when we place these elements in the compounds considered here, each atom of these species has one 5$d$ electron, thereby restoring the sequence. As expected the 4$f$ electrons are polarised and the 5$d$ are not. The spin polarisation $N_{up}-N_{down}$ is therefore 1, 2 and 3 electrons for the Ce, Pr and Nd atoms in their respective compounds. The 5$d$ electron is unpolarised and strongly hybridised with other bands (Figure \ref{BiSbands}).

These compounds share the space group P4/nmm. The lanthanide atoms are on the sites (1/4,1/4,$z$) and (3/4,3/4,-$z$). These have a 4mm symmetry. Hence the sites have the point group $C_{4v}$. The nearest neighbours of the lanthanide atoms are 4 oxygen atoms. As a way of characterising the 4$f$ orbitals in this environment, we treat the \textit{Ln} and the 4 O atoms as a molecule and look for the symmetries of the molecular orbitals for bonding in that molecule. Finally, we obtain that all the 4$f$ orbitals are allowed by symmetry to hybridise with the oxygen atoms, except the 4$f_{zx^2 - zy^2}$ orbital, but it is never occupied, see below.

\subsubsection{CeO$_{0.5}$F$_{0.5}$BiS$_{2}$}

In the cerium compound we find two different competing states which are quasi-degenerate in energy. It does not make sense for us to try to establish which one is favoured since either state appears in the calculations under very slight changes of the technical parameters, even for very tightly converged settings (see the Methods section).

In the ferromagnetic configuration, calculations always converge on the same state which we call state F. But in the antiferromagnetic configuration, calculations converge on two different states, state AF1 and state AF2. They differ by the 4$f$ orbital chosen by the 4$f$ electrons in the Ce atoms. The orbital occupations of the 4$f$ electron are summarised in Table \ref{OrbitalsCe}. The total energy of state AF2 is higher than the one of state F, while the total energy of state AF1 is lower than the one of state F (Table \ref{EnergiesCe}). Therefore we obtain two competing states, F and AF1.

\begin{table}[h]
\center{
	\begin{tabular}{ccc}
		Configuration & \multicolumn{2}{c}{4$f$ occupations}\\
		& 1st Ce atom & 2nd Ce atom \\
		\hline
		F   & 4$f_{5yz^2 - yr^2}$ & 4$f_{5yz^2 - xr^2}$ \\
		AF1 & 4$f_{5yz^2 - yr^2}$ & 4$f_{5yz^2 - xr^2}$ \\
		AF2 & 4$f_{y^3 - 3x^2y}$ & 4$f_{5yz^2 - xr^2}$ \\
		\hline
	\end{tabular}
	\caption{Orbitals occupied by the 4$f$ electron in each cerium atom in the unit cell of CeO$_{0.5}$F$_{0.5}$BiS$_{2}$ in the two antiferromagnetic configurations AF1 and AF2. The labelling order of the cerium atoms is arbitrary.
	}
	\label{OrbitalsCe}
}
\end{table}

\begin{table}[h]
\center{
	\begin{tabular}{cc}
		Configuration & Total energy (eV)\\
		\hline
		F   & 0\\
		AF1 & -0.0124\\
		AF2 & 0.1105\\
		\hline
	\end{tabular}
	\caption{Total energies of the various magnetic configurations obtained for CeO$_{0.5}$F$_{0.5}$BiS$_{2}$ with respect to the energy of F.
	}
	\label{EnergiesCe}
}
\end{table}

In state AF1 the spin-spin interaction is weak, with a slight tendency for them to order antiferromagnetically, as manifested by the energy difference with respect to ferromagnetic order of 12 meV per cell. In state F the ferromagnetic order is more substantially favoured, by 111 meV per cell.

\begin{figure}
\center{\includegraphics[scale=0.3]{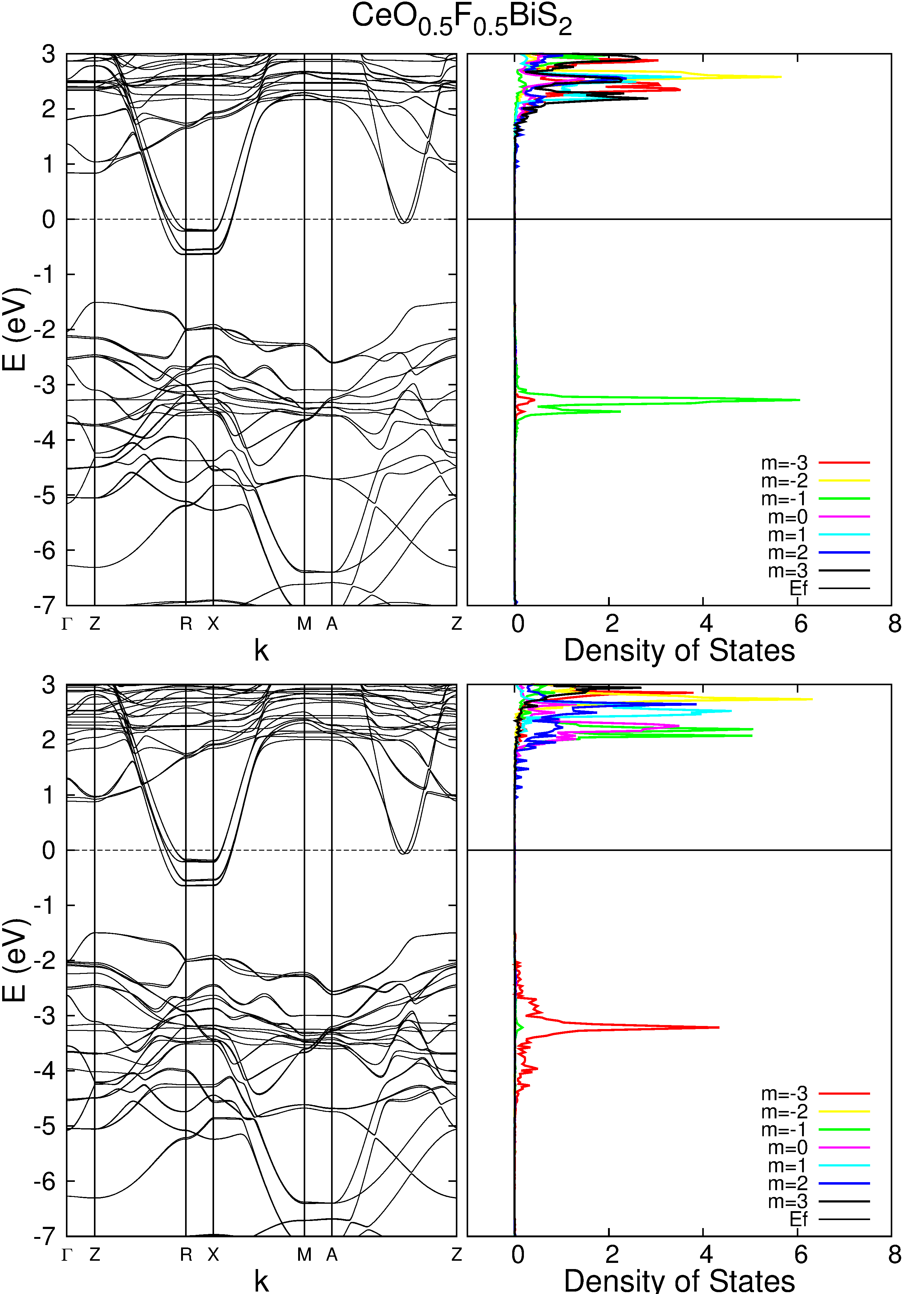}
\caption{Band structures and density of 4$f$ states of CeO$_{0.5}$F$_{0.5}$BiS$_{2}$ in the antiferromagnetic configuration, in the two states that it converges to. The projected densities of states were plotted using a peak width of $0.01$ eV, smaller than the standard $0.2$ eV used in the rest of this work. State AF1 is at the top, and state AF2 at the bottom. The decomposition of density of states on the 9 orbitals clearly shows that the two configurations differ by the jump of one electron from one orbital to another.}
\label{Ce-AFM-comparison}}
\end{figure}

The band structures and density of states for states AF1 and AF2 are plotted in Figure \ref{Ce-AFM-comparison}. The bands are very similar, and quite indistinguishable around the Fermi level. The only difference is that in state AF2 the bands around the peak in $4f$ density of states are slightly higher than in state AF1. However, the projected density of states on the different $f$-orbitals clearly show the change of orbital character of the 4$f$ electron.

The ferromagnetic state is compatible with experimental observations of ferromagnetism in this compound at low temperature. The presence of the weak antiferromagnetic state suggests that this system may be close to an antiferromagnetic instability associated to a change of orbital for the 4$f$ electron, which would suggest possibly interesting phenomena associated with this coupling between spin ordering and choice of orbital.

\subsubsection{PrO$_{0.5}$F$_{0.5}$BiS$_{2}$}

A similar instability is observed in the Pr compound, in which each praseodymium atom has two electrons in the 4$f$ shell. We obtain an effective degeneracy between two states for both the ferromagnetic (states F1 and F2) and the antiferromagnetic (states AF1 and AF2) configurations. In all cases the spin polarisation is 2 electrons on each atom. The two states on which each configuration converges only differ by the orbital occupations of the 4$f$ electrons of the praseodymium atoms. The orbital occupations of the 4$f$ electron are summarised in Table \ref{OrbitalsPr}.

\begin{table}[h]
\center{
	
	\begin{tabular}{ccccc}
		Configuration & \multicolumn{4}{c}{4$f$ occupations}\\
		& \multicolumn{2}{c}{1st Pr atom} & \multicolumn{2}{c}{2nd Pr atom} \\
		& 1st electron & 2nd electron & 1st electron & 2nd electron \\
		\hline
		F1 & 4$f_{5yz^2 - yr^2}$ & 4$f_{5yz^2 - xr^2}$ and 4$f_{x^3 - 3xy^2}$ &  4$f_{5yz^2 - xr^2}$ & 4$f_{5yz^2 - yr^2}$ and 4$f_{y^3 - 3x^2y}$ \\
		F2 & 4$f_{xyz}$ & 4$f_{5yz^2 - yr^2}$ and 4$f_{y^3 - 3x^2y}$ & 4$f_{5yz^2 - xr^2}$ & 4$f_{5yz^2 - yr^2}$ and 4$f_{y^3 - 3x^2y}$ \\
		AF1 & 4$f_{xyz}$ & 4$f_{5yz^2 - xr^2}$ and 4$f_{x^3 - 3xy^2}$ & 4$f_{xyz}$ & 4$f_{5yz^2 - xr^2}$ and 4$f_{x^3 - 3xy^2}$\\
		AF2 & 4$f_{xyz}$ & 4$f_{5yz^2 - xr^2}$ and 4$f_{x^3 - 3xy^2}$ & 4$f_{5yz^2 - xr^2}$ & 4$f_{5yz^2 - yr^2}$ and 4$f_{y^3 - 3x^2y}$\\
		\hline
	\end{tabular}
	\caption{Orbitals occupied by the 4$f$ electron in each praseodymium atom in the unit cell of PrO$_{0.5}$F$_{0.5}$BiS$_{2}$ in the four configurations FM1, FM2, AF1 and AF2. The labelling order of the praseodymium atoms and electrons is arbitrary.
	}
	\label{OrbitalsPr}
}
\end{table}

\begin{table}[h]
\center{
	\begin{tabular}{cc}
		Configuration & Total energy (eV)\\
		\hline
		F1  & 0\\
		F2  & 0.094527\\
		AF1 & 0.092904\\
		AF2 & -0.000248\\
		\hline
	\end{tabular}
	\caption{Total energies of the various magnetic configurations obtained for PrO$_{0.5}$F$_{0.5}$BiS$_{2}$ with respect to the total energy of F1.
	}
	\label{EnergiesPr}
}
\end{table}

For the same input parameters we obtain either F1 and AF1 or F2 and AF2. Therefore for a given set of parameters we obtain two energy differences, AF1-F1 and AF2-F2. These are $92.9$ meV in favor of ferromagnetism and $94.8$ meV in favor of antiferromagnetism, hence we obtain two competing states, AF2 and F1 (Table \ref{EnergiesPr}).

The Pr compound has therefore two competing states, with very close energies. Both display magnetic ordering, with a strength very close to the one in CeO$_{0.5}$F$_{0.5}$BiS$_{2}$, one ferromagnetic and the other antiferromagnetic.

\subsubsection{NdO$_{0.5}$F$_{0.5}$BiS$_{2}$}

In this case we find a single ferromagnetic state and a single antiferromagnetic one. We obtain a difference between the two spin states of 3.3 meV favouring antiferromagnetic order. This is very small, indeed it is two orders of magnitude lower than the difference in favor of ferromagnetism in the cerium compound. It is therefore likely that the magnetic order in this material will arise at temperatures much lower than the Curie temperature of the order of 1 K measured in the cerium compound.

The presence of an instability in the cerium and praseodymium doped compounds is consistent with the well known valence fluctuation phenomenon which has been studied in other compounds containing these two atoms \cite{Holmes2007,Herrero-Martin2011}. To the knowledge of the authors, no neodymium compound exhibits this phenomenon, which is in line with its absence in NdO$_{0.5}$F$_{0.5}$BiS$_{2}$.

The results discussed above for the three compounds do not change qualitatively with the choice of $U$ value, as shown in Figure \ref{MagnetismU}, where the energy difference between the antiferromagnetic and the ferromagnetic configurations are shown for the three compounds. Calculating the total energy of other magnetic configurations using a larger unit cell has been attempted but definite conclusions could not be achieved, partly because of the described instabilities.

\subsection{CeOBiS$_{2}$ under pressure}

\begin{figure}
\center{\includegraphics[scale=0.3]{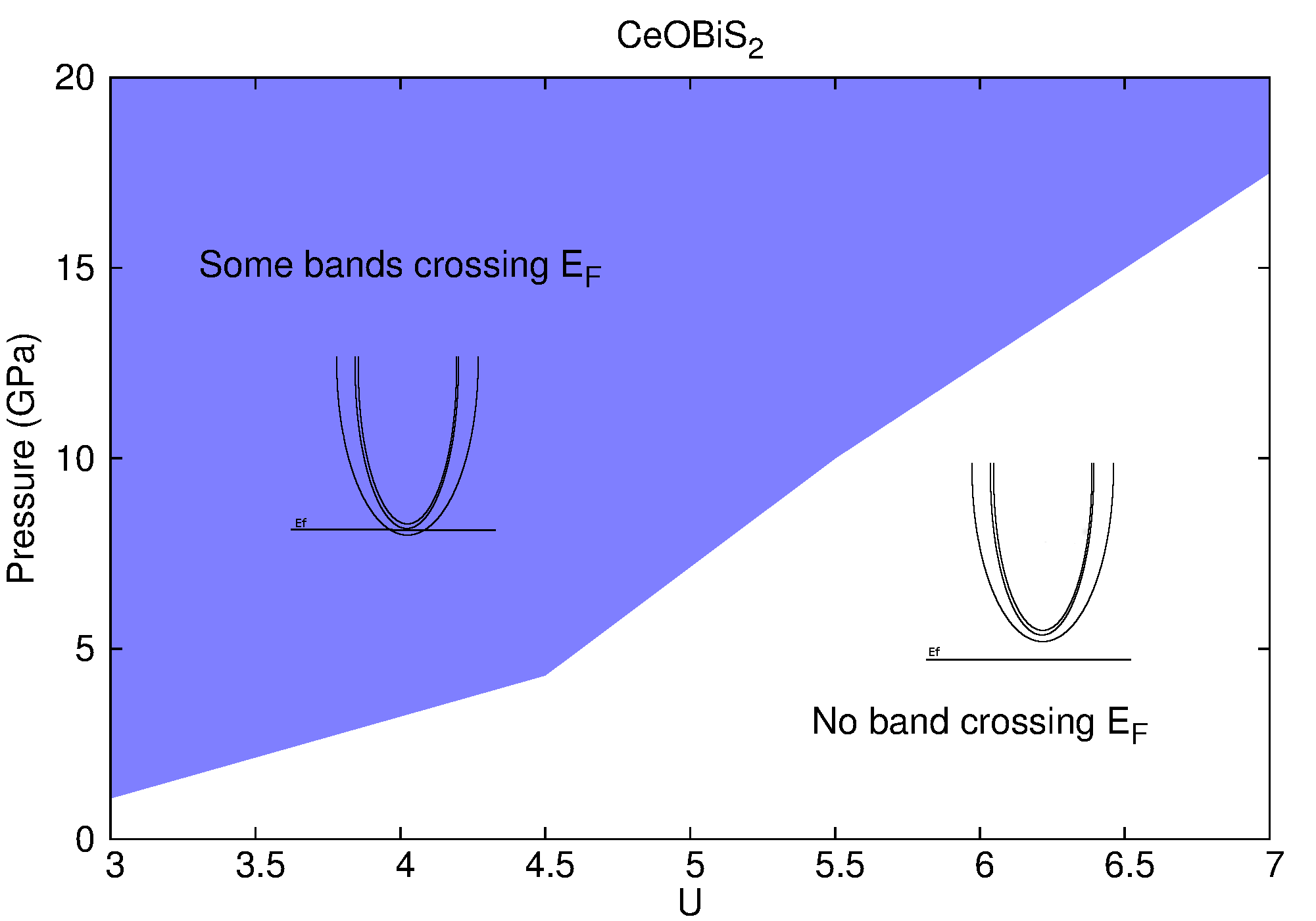}
\caption{Pressure / $U$ phase diagram of CeOBiS$_{2}$.}
\label{CeOBiS2}}
\end{figure}

\begin{figure}
\center{\includegraphics[scale=0.5]{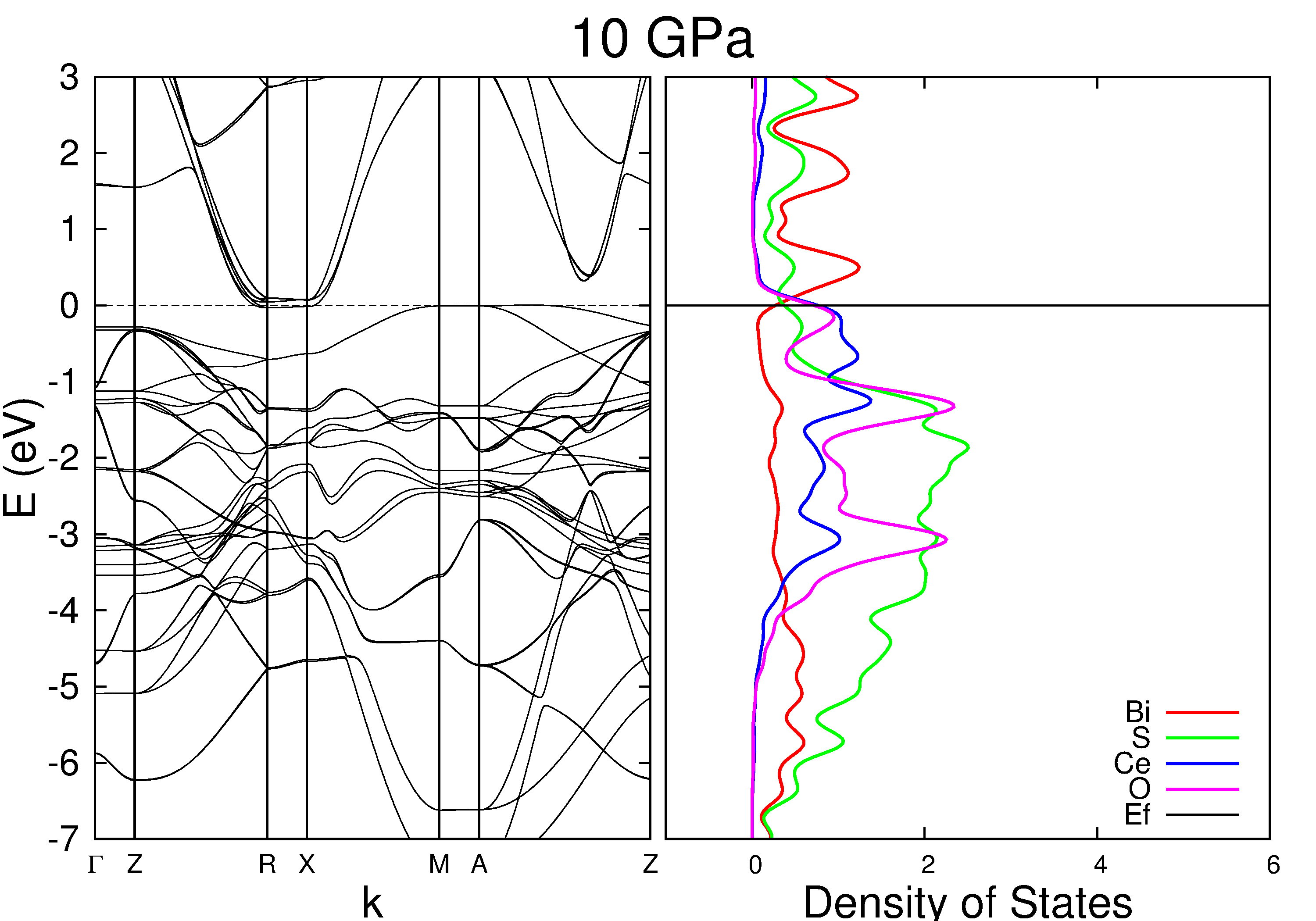}
\caption{Band structure (left) and density of states projected onto the basis orbitals for Bi (red), S (green), Ce (blue), and O (lilac) (right) for CeOBiS$_2$ under $10$ GPa. We use $U=5$ and spin polarisation.}
\label{CePressure}}
\end{figure}

Finally, we consider the effect of pressure on CeOBiS$_{2}$. Without electron doping the system is insulating with the BiS bands clearly above the Fermi level. The question is whether pressure can close that gap. The band structure of CeOBiS$_{2}$ has been calculated with respect to pressure and $U$ (Figure \ref{CeOBiS2} and Figure \ref{CePressure}). At ambient pressure, the system is insulating. When pressure increases, the gap closes and some band cross the Fermi level. The transition pressure varies with $U$, from 1.2 GPa for $U=3$ eV to 17.5 GPa for $U=7$ eV. The bands crossing the Fermi level are not only the BiS bands: other bands are crossing. These are cerium bands, hybridised with oxygen and sulphur bands. Indeed, the main effect of pressure is to raise the cerium oxide bands which in turn dope the BiS bands. This effect is similar to the stoichiometric doping of the BiS bands by the sulphur dimers bands in Bi$_{3}$O$_{2}$S$_{3}$ \cite{Morice2015}. The contribution of the oxygen and sulphur bands decreases with pressure but remains substantial, as can be seen in the density of states. The composition of the BiS bands does not change with pressure, it is always dominated by the same Bi 6$p$ and S 3$p$ states. Finally, the conduction band is narrowed whereas the valence band is slightly widened with pressure. Detailed plots of the band structures and densities of states, and crystal structures obtained are available in the supplementary material. 

The insulating state at ambient pressure is consistent with the fact that no superconductivity has been observed in this compound. It shows that there should be a insulator-metal transition under pressure, which also could be superconducting at low temperature. The transition pressures obtained have rather large uncertainties given that they have been obtained through DFT-GGA (with its well-known bad gap problem), but we can still conclude that it seems likely the transition temperature could be reachable by commonly known techniques. Of course, this pressure also depends heavily on the $U$ parameter relevant for the compound.

The type of ``doping" involved is different from the fluorine type. Indeed, here pressure brings the valence and the conduction bands together until they cross. Pressure has opposite effects on the valence and conduction bands: the conduction band is narrowed and its composition does not change, whereas the valence band is widened and its hybridisation with oxygen and sulphur orbitals is decreased.

\section{Conclusion}

We calculated the band structure of the four doped compounds and showed that the change of rare earth does not affect the Fermi surface. We calculated the local density of states, which clearly indicates that the conduction electrons are confined to the bismuth-sulphur planes. 
 
We calculated the difference in energy between the two possible magnetic configurations one can obtain in one unit cell of \textit{Ln}O$_{1-x}$F$_{x}$BiS$_{2}$ (\textit{Ln} = Ce, Pr, and Nd). We demonstrated that CeO$_{0.5}$F$_{0.5}$BiS$_{2}$ has competing ferromagnetic and weak antiferromagnetic tendencies. The first case corresponds to experimental results. PrO$_{0.5}$F$_{0.5}$BiS$_{2}$ has a strong tendency to order, which could be ferromagnetically or antiferromagnetically depending on subtle differences in 4$f$ orbital occupations. NdO$_{0.5}$F$_{0.5}$BiS$_{2}$ has a stable weakly antiferromagnetic competing state.

Finally, we did predict that applying pressure to CeOBiS$_{2}$ would make it metallic, and possibly superconducting, through raising the cerium oxide bands. This type of doping of the BiS bands has not been investigated in the BiS$_2$ family so far and could have novel properties.

\section*{Acknowledgements}

We would like to thank Arman Khojakhmetov for his exploratory work, and Gilbert G. Lonzarich, Peter Littlewood, Pablo Aguado Puente, Stephen Rowley, Sebastian Haines, Cheng Liu, Daniel Molnar and Adrien Amigues for fruitful discussions. We acknowledge computing resources of the Spanish Supercomputer Network (RES), and support from EPSRC, Corpus Christi College and the  Winton
Programme for the Physics of Sustainability.

\section*{References}

\bibliographystyle{unsrt}
\bibliography{BiS2-based,DFT,Layered-materials}

\end{document}